# A graph-based approach for modification site assignment in proteomics

*Dafni Skiadopoulou[1,2], Lukas Käll[3], Harald Barsnes[2,4], Veit Schwämmle[5], Marc Vaudel[1,2,6,$]*

[1] Mohn Center for Diabetes Precision Medicine,
  Department of Clinical Science,
  University of Bergen, Bergen, Norway

[2] Computational Biology Unit,
  Department of Informatics,
  University of Bergen, Bergen, Norway

[3] Science for Life Laboratory,
  Department of Gene Technology,
  KTH - Royal Institute of Technology, Stockholm, Sweden

[4] Proteomics Unit,
  Department of Biomedicine,
  University of Bergen, Bergen, Norway

[5] Department of Biochemistry and Molecular Biology,
  University of Southern Denmark, Odense, Denmark.

[6] Department of Genetics and Bioinformatics,
  Norwegian Institute of Public Health, Bergen, Norway

[$] To whom correspondence should be addressed

## Abstract

*Background*
In proteomics, the most probable localizations of post-translational modifications are assessed by localization scores evaluating the likelihood of a given modification to occupy a site on a peptide sequence. When identifying highly modified peptides, localization scores for different modifications can return conflicting results, stacking modifications on the same amino acid. Here, we propose a graph-based approach that assigns modifications to sites in a way that maximizes localization scores while avoiding conflicting assignments.

*Results*
The algorithm is implemented as both a standalone Python program and in the compomics-utilities Java library. Our graph-based approach showed the ability to match complex combinations of modifications and acceptor sites, allowing the processing of thousands of peptides in a few seconds.

*Conclusions*
Our graph-based approach to modification site assignment allows distributing multiple modifications in a way that maximizes individual localization scores. Having an optimal modification site assignment is important for spectrum annotation and biological interpretation.

## Abbreviations
PTM: post-translational modification
PSM: peptide-spectrum match

## Background
In mass spectrometry-based proteomics, mass spectra of fragmented peptides are matched against a database of protein sequences using a proteomic search engine, allowing the high-throughput identification of peptide sequences together with their modification status (1). The localization of post-translational modifications (PTMs) on protein sequences can provide important functional information (2). Conversely, incorrect PTM localization reduces the quality of spectrum annotation, yielding lower identification scores, and possibly erroneous biological inference (3). To evaluate the localization of modifications on a peptide sequence, multiple localization scores were developed that estimate the likelihood for a given acceptor site to be occupied for every modification of every peptide-spectrum match (PSM) (4,5).

Once the localization scores have been computed, the modifications are assigned to the sites of highest score before further processing, e.g. error rate estimation using Percolator (6). It is important to note that due to differences in scoring, and especially since search engines often only consider the most intense peaks in mass spectra, the peptide maximizing the modification localization scores is not necessarily the best scoring peptide in the search engine results. And since the search engines also often have limitations in terms of how many combinations of modification sites to consider, the peptide maximizing localization scores might not even be in the list of peptide candidates returned by the search engine (7). An agnostic rescoring considering all possible acceptor sites is therefore necessary.

In order to find the peptide maximizing the modification localization scores, the modifications found by the search engine need to be assigned to the site with highest localization score following assignment rules: modifications have different types of acceptor sites, e.g. amino acids or termini, and should not be stacked on the same acceptor site. While maximizing the scores is trivial with a couple of modifications, e.g. phosphorylation and oxidation, the problem becomes more complex when combining several modifications with possible conflicting sites. Furthermore, since this needs to be conducted on millions of PSMs per experiment, the site assignment needs to be computationally fast. While PTM-Shepherd addresses this problem for open modification searches (8), to the best of our knowledge no method has been designed to find the modification configuration maximizing localization scores.

Here, we propose a graph-based approach that models modifications and their acceptor sites, and returns a configuration that maximizes localization scores with controlled processing time. The modification assignment problem is reduced to the maximum weight maximum cardinality matching problem in bipartite graphs, by modelling modifications and acceptor sites as vertices, connected by edges whose weights are calculated based on the corresponding modification localization scores. We provide a standalone Python implementation as well as a Java implementation in the compomics-utilities library (9) and demonstrate its usage in PeptideShaker (10).

## Implementation

*Model*

In this work, the modification site localization problem is addressed by a graph approach. For each PSM, a weighted bipartite graph ($G = \{V, E, w\}$) is used to model all possible combinations of modifications on the amino acid chain. In this graph we determine two kinds of vertices, the ones that represent the different modifications (i.e. $D = \{d_1, d_2, \ldots, d_k\}$) and the ones that represent all their possible acceptor sites (i.e. $A = \{a_1, a_2, \ldots, a_n\}$). The set of vertices V is then formed by the union of the sets A and D (i.e. $V = A \cup D$). For example, a peptide carrying four acetylations and one dimethylation, and featuring eight acceptor sites on the amino acid chain where either modification can occur, is modeled by a graph with 13 vertices in total ( $/\!/ D /\!/ = 5$, $/\!/ A /\!/ = 8$, hence $/\!/ V /\!/ = 13$). For each modification in D, an edge is formed between its vertex and the vertex of each of its possible acceptor sites in A. A weight $w_{i,j}$ is assigned to each edge $e_{i,j}$ corresponding to the localization score of the modification $d_i$ to the acceptor site $a_j$.

Using this model, we reduce the modification site assignment problem to the maximum weight maximum cardinality matching problem on the resulting bipartite graph. The solution of this problem will be a set of pairwise non-adjacent edges satisfying two conditions: (i) the matching will be of maximum cardinality, which in this modeling ensures that all the vertices that represent a modification will be matched; and (ii) the weight of the matching is maximum, which means that it provides the combination of modifications and acceptor sites that maximizes the sum of localization scores.

*Implementation*

We provide an open source Python-based implementation of our approach available under a GPL-3.0 license (github.com/ProGenNo/peptides-modifications-matching). For this, we used the networkX (11) library and its algorithm for the minimum weight full matching problem. This algorithm when applied on our model produces a matching M with cardinality $/\!/ M /\!/ = \min\{/\!/ A /\!/, /\!/ D /\!/\} = /\!/ D /\!/$ which ensures that all the modifications will be matched to a site in the amino acid chain (because the model assumes that the graph is connected and there are at least as many possible acceptor sites as there are modifications, therefore $/\!/ A /\!/ \geq /\!/ D /\!/$). However, the algorithm produces a matching that minimizes the sum of weights of the corresponding edges. Therefore, in order to be suitable for the modification assignment problem, the weights of the edges from the model described in this work were transformed accordingly so as the maximum localization score corresponded to the minimum edge weight. The implementation script receives a text file as input, where a description of the available PSMs is included. It requires one line per PSM with the information of the spectrum and peptide IDs as well as the modifications with their possible acceptor sites and the corresponding localization scores. Exceptions are thrown for each PSM if there are fewer acceptor sites given than modifications, or if there is no possible full matching in the resulting graph, or if the localization scores take negative values. Further information is available at the GitHub repository.

We also provide an open source Java-based implementation freely available as part of the compomics-utilities library (9) under an Apache-2.0 license. It is located in the com.compomics.util.experiment.identification.modification.peptide_mapping package and makes use of the JGraphT package (12). This includes an implementation of the Kuhn-Munkres algorithm for the assignment problem (13), which requires as input a complete bipartite graph and produces a perfect matching with minimum weight. To make this algorithm suitable for the modification assignment problem the weights of the edges of the model proposed in this work were transformed as in the Python implementation. Also, the graph model was made complete with the addition of dummy vertices and edges if needed, while ensuring that the perfect matching of minimum weight still corresponded to the best assignment of modifications to acceptor sites. The implementation in compomics-utilities has been integrated in the processing of search engine output with PeptideShaker (10), enabling its use without requiring programming skills.

*Performance benchmark*

We generated sets of random peptide sequences of length 30 with multiple random sets of modifications. For each peptide, one to ten distinct modifications were randomly generated. For every type of modification, up to six modification sites were randomly selected, and the number of modified residues was randomly set between one and the number of modification sites. Finally, modification localization scores were given randomly for all modification sites, and the matching of modification to site was conducted using the Java implementation of the compomics-utilities library. The code was run for different batches of peptides using different numbers of threads on a standard laptop computer (Lenovo ThinkPad P15 Gen 2i from 2021, 11th Gen Intel® Core™ i9-11950H × 16, 64 GB of RAM, operated using Ubuntu 24.04.2 LTS), and the processing time recorded.

Subsequently, we generated peptides in a similar fashion, but fixed the number of distinct modifications, modification sites, and modified residues. We then benchmarked the processing time for all possible combinations using 1,000 peptides per thread and all threads available.

Each benchmark was replicated ten times. The code used to benchmark the application is available in the

com.compomics.util.experiment.identification.modification.peptide_mapping.performance package of the compomics-utilities library.

*Example*

A set of ten raw files from a publicly available dataset of mouse histones (14) obtained from ProteomeXchange (15), identifier PXD005300, via the PRIDE partner repository (16) was searched using MS-GF+ (17) using SearchGUI (18). No fixed modification was selected, multiple variable modifications were selected to create complex site assignment combinations: Acetylation (K, N-terminus), Ammonia loss (N), Carbamidomethylation (C), Deamidation (N, Q), Dimethylation (K), Dioxydation (M), Methylation (K), Oxidation (K, M), Phosphorylation (S, T, Y), Trimethylation (K). Fragment ions c and z were accounted for using a tolerance of 1.05 Da and 10 ppm for MS1 and MS2, respectively. Enzyme was set to Glu-C, allowing up to 30 missed cleavages, 10 Da isotopic error and 100 amino acid lengths to avoid excluding long peptides with many acceptor sites. The result file was processed using PeptideShaker where all modifications were scored using PhosphoRS (19) and sites were assigned using the implementation from the compomics-utilities library.

## Results and Discussion

We designed a graph-based approach to the problem of assigning PTMs to acceptor sites in a way that maximizes localization scores while avoiding stacking multiple modifications on the same acceptor site. We provide implementations in Python and Java for scripting applications and usage in bioinformatic tools, and demonstrate its usage in PeptideShaker for search engine result post-processing.

*Example on a histone data set*

To demonstrate an application of the approach, we searched a set of publicly available mass spectrometry files obtained on mouse histones using multiple variable modifications, where several modifications can compete for the same sites. On a standard laptop, the database search took approximately 3.5 hours, the scoring of the modification localization took a minute, and the modification site assignment took less than ten seconds. Two examples of graph models for multiply modified peptides are presented in Figure 1 along with the site assignment selected by the score maximization algorithm. More examples are available in our GitHub repository (github.com/ProGenNo/peptides-modifications-matching). The maximum weight matching assigned the modifications to the best-scoring residues while avoiding stacking modifications. If multiple combinations give the same total sum of localization scores, one of them will be chosen at random. Note that this study solely focuses on the ability of the algorithm to match PTMs to acceptor sites and the results of the peptide identification are outside the scope of this paper.

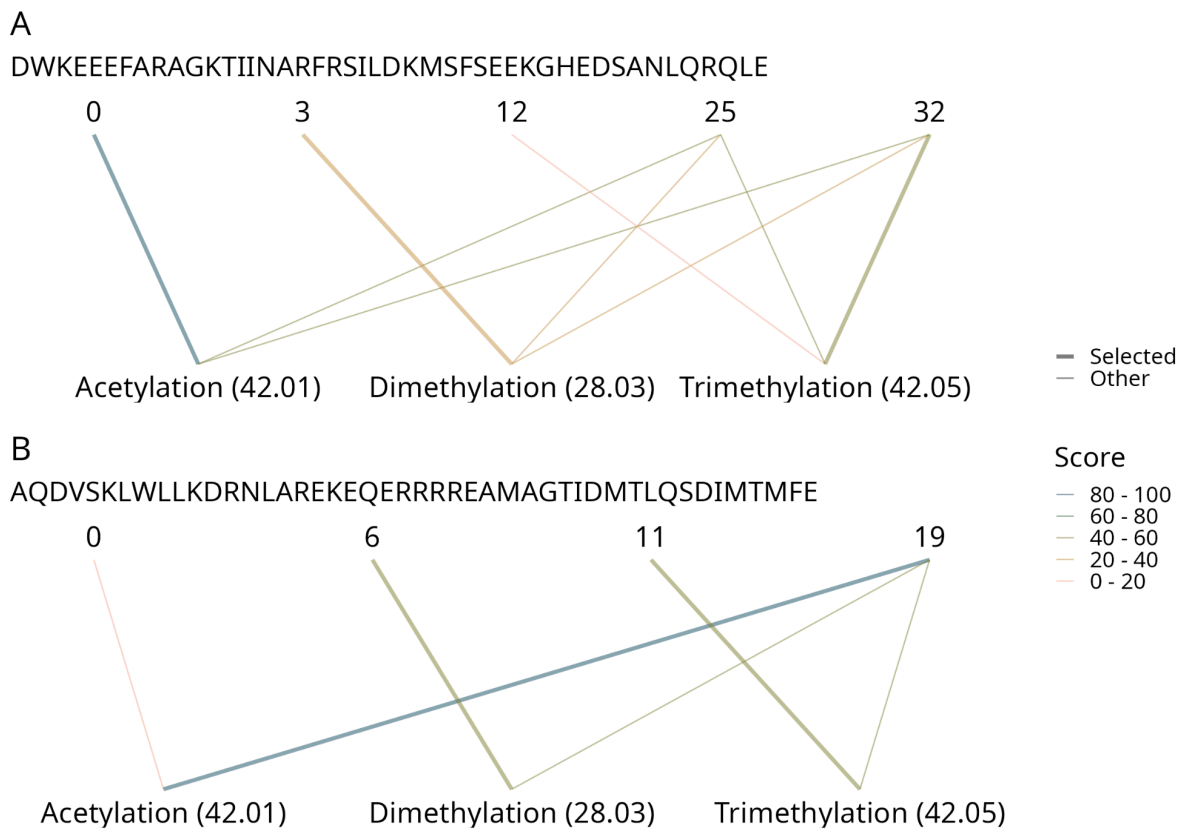

*Figure 1: Examples of modification assignment for two sets of modification localization scores. The sequence of the peptide is presented at the top with acceptor sites for modifications represented as their 1-based index on the sequence, with zero and peptide length + 1 representing the N- and C-terminus, respectively. The modifications with their rounded mass are linked to the sites with an edge which color indicates the localization score, and the edge is represented as a thick or thin line if the site was selected by the matching approach or not, respectively.*

*Performance*

We benchmarked the performance of the matching approach using synthetically generated peptides (see *Performance benchmark* for details). Around one million peptides could be processed per minute per thread. Since the peptides are scored individually, they can be processed in parallel and as soon as approximately 1,000 peptides were available, the processing time increased linearly with the number of peptides and a minor increase in processing time was observed when using multiple threads, which we impute to Java having to manage considerably more objects (Figure 2A).

For a given peptide, the modification matching time increased with the number of distinct modifications, the number of sites, and the number of modified residues (Figure 2B). In all configurations, the modification site assignment took under two seconds for thousands of peptides, which is negligible compared to the search time, which can be several hours.

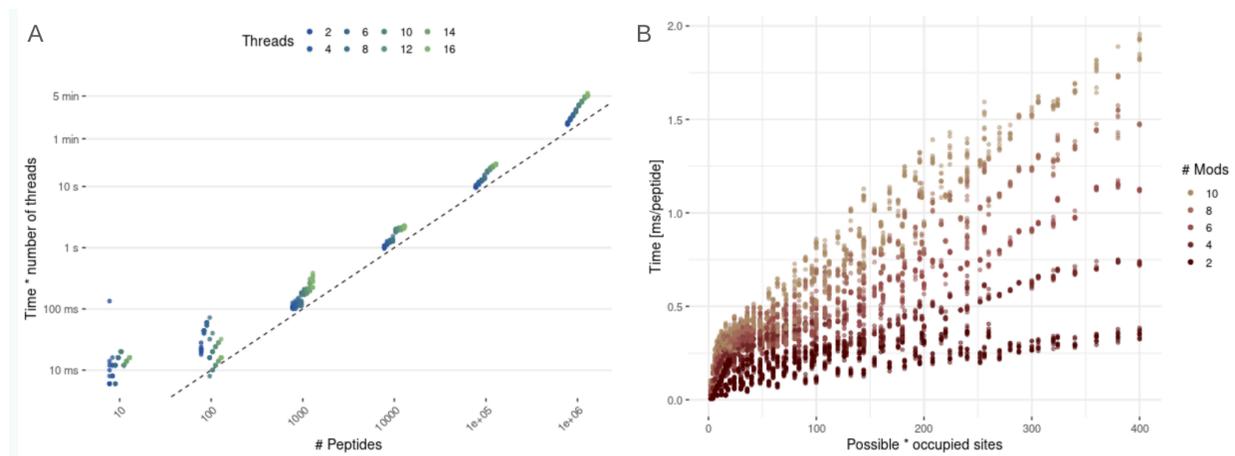

*Figure 2: Performance on simulated peptides. (A) Computation time per thread for different sets of peptides on a log-log scale. (B) For a given set of peptides, the processing time for different numbers of modifications against the number of possible sites multiplied by the number of occupied sites.*

*Limitations*

By design, the implementation does not decompose or stack modifications, e.g. transform a dimethylation into two methylations, or vice versa. It relies on the search engine returning at least one PSM with the correct set of modifications. Addressing this limitation, e.g. in the case of single, di-, and trimethylation would require considering all possible PTM decompositions and combinations. This is compatible with the approach proposed here but requires extending the post-search PTM handling and scoring.

The method relies on the hypothesis that the localization score of a PTM remains reliable after the relocalization of another PTM. This can be violated in case a change in localization of a PTM affects the site-determining ions of another PTM. This problem stems in the computational and mathematical difficulty of scoring the localization of multiple PTMs at the same time. It can be alleviated through iterative scoring, but for the confident localization of highly modified peptides, better localization scores are needed that can take multiple modifications into account.

## Conclusions

This work addresses the problem of finding the optimal assignment of PTMs on the amino acid chain of peptides upon modification localization scoring in mass spectrometry-based proteomics. Given the possible modifications and all of their acceptor sites together with the corresponding localization scores, the proposed graph-based approach finds the combination of modifications on the amino acid chain that maximizes the localization scores while avoiding impossible combinations. This approach thus enables the consolidation of modification localization scores for heavily modified peptides and thereby increases the performance of proteomic pipelines when multiple modifications are considered without substantially increasing processing time.

## Availability and requirements

Project name: peptides-modifications-matching
Project home page: https://github.com/ProGenNo/peptides-modifications-matching
Operating system(s): Platform independent
Programming language: Python and Java
License: GPL-3.0 (Python implementation) and Apache 2.0 (Java implementation)

## Declaration

*Availability of data and materials*
The histone data set is available from ProteomeXChange via the PRIDE partner repository with accession PXD005300.

*Authors' contributions*
DS and MV designed the project, implemented the approach, benchmarked the results, and wrote the paper. VS contributed ideas, expertise on PTM analysis, and wrote the paper. HB and LK contributed to writing the paper.

*Acknowledgements*

*Competing interests*
The authors declare that they have no competing interests.

*Funding*
This work was supported by the Research Council of Norway (project #301178 to MV) and the University of Bergen.

This research was funded, in whole or in part, by the Research Council of Norway 301178. LK was funded by the Swedish Research Council (Grant 2024-05887). A CC BY or equivalent license is applied to any Author Accepted Manuscript (AAM) version arising from this submission, in accordance with the grant's open access conditions.

# Background